\documentstyle[prl,aps,epsf,multicol,tabularx]{revtex}
\begin{document}

\title{The role of diffusion in branching and annihilation random walk
models}
\author{G\'eza \'Odor}
\address{Research Institute for Technical Physics and Materials Science, \\
H-1525 Budapest, P.O.Box 49, Hungary}    
\maketitle

\begin{abstract}
Different branching and annihilating random walk models are investigated by 
cluster mean-field method and simulations in one and two dimensions. 
In case of the $A\to 2A$, $2A\to\emptyset$ model the cluster mean-field 
approximations show diffusion dependence in the phase diagram as was found 
recently by non-perturbative renormalization group method  
(L. Canet et al., cond-mat/0403423). 
The same type of survey for the $A\to 2A$, $4A\to\emptyset$ model results 
in a reentrant phase diagram, similar to that of $2A\to 3A$, $4A\to\emptyset$ 
model (G. \'Odor, PRE {\bf 69}, 036112 (2004)).
Simulations of the $A\to 2A$, $4A\to\emptyset$ model in one and two 
dimensions confirm the presence of both the directed percolation 
transitions at finite branching rates and the mean-field transition at zero 
branching rate. In two dimensions the directed percolation transition 
disappears for strong diffusion rates. These results disagree with the 
predictions of the perturbative renormalization group method.
\end{abstract}
\begin{multicols}{2}

\section{Introduction}

Phase transitions in nonequilibrium, dynamical systems, which do not satisfy 
the detailed balance condition and do not possess hermitian Hamiltonian may 
appear in models of population, epidemics, catalysis, cooperative 
transport \cite{DickMar}, enzyme biology \cite{Berry} and markets for 
example \cite{Bou}.
Many of the known systems can be mapped onto some reaction-diffusion
models, the behavior of them are the studied intensively in the past decades
\cite{DickMar,Hin2000}. In these systems particle (A) creation, 
annihilation and diffusion processes compete and by tuning the control 
parameters phase transition may occur from an active steady state to an 
inactive, absorbing state of zero density. The simplest example of such
models exhibiting phase transition are the branching and annihilating random 
walk models (BARW), in which offsprings are created by a single ancestor:
$A\to(n+1)A$ and particles annihilate: $2A\to\emptyset$.

The classification of universality classes of nonequilibrium systems is one
of the most important tasks of statistical physics \cite{Uweof,dok}.
Universal scaling behavior may occur at continuous phase transitions like in 
equilibrium systems and the corresponding $n$-point correlations are 
homogeneous (but anisotropic) functions of space and time. 
In the past decades analytical and numerical studies explored a large 
zoo of such classes \cite{dok}.
One may pose the important question: which are the relevant factors 
determining these classes ?
By inspecting the widespread literature it appears that besides the well 
known factors of homogeneous equilibrium models with short ranged interactions:
i.e. spatial dimension, symmetries, conservation laws there are more
things must be taken into account. For example due to the possibility of
transitions in low dimensions topological effects may play important
role \cite{mexprocikk}. Furthermore initial condition -- as a boundary in 
the time direction -- can also be relevant \cite{PCP,HaOd98}.

The mean-field classes of general,
\begin{equation}
n A \stackrel{\sigma}{\to} (n+k)A,
\qquad m A \stackrel{\lambda}{\to} (m-l) A,
\qquad\emptyset A\stackrel{D}{\leftrightarrow} A\emptyset
\label{genreactions}
\end{equation}
site occupation number restricted models 
(with $n>1$, $m>1$, $k>0$, $l>0$ and $m-l\ge 0$) 
resulted in a series of different universality classes 
depending on $n$ and $m$ \cite{tripcikk}. This means that 
above the upper critical dimension ($d_c$) of the transition
{\it $n$ and $m$ are relevant parameters}. 
In particular for the $n=m$ symmetrical case the density of particles
in the active phase, near the critical point ($\sigma_c > 0$) scales as
\begin{equation}
\rho \propto |\sigma-\sigma_c|^{\beta}, \label{betascale}
\end{equation}
with $\beta=1$, while at the critical point it decays as
\begin{equation}
\rho \propto t^{-\alpha} \ , \label{alphascale}
\end{equation}
with $\alpha=\beta/\nu_{||}=1/n$ \cite{PHK02,tripcikk}.
On the other hand for the $n<m$ asymmetric case continuous phase 
transitions at zero branching rate $\sigma_c=0$ occur characterized by
\begin{equation}
\beta=1/(m-n), \quad \alpha=1/(m-1) \label{asymmfscal}
\end{equation}
For $n>m$ the mean-field solution provides first order
transition.

By going beyond the site mean-field approximation it turned out that the
phase diagrams of the type eq.(\ref{genreactions}) models may contain other
transition points with non-trivial scaling behavior.
In previous papers \cite{2340cikk,bin2dcikk} I investigated the $2A\to 3A$, 
$4A\to\emptyset$ model by cluster mean-field approximations and simulations 
in one and two dimensions and showed that the {\it diffusion} plays an 
important role: it introduces a different critical point besides the one 
obtained by site mean-field solution at $\sigma_c=0$ (\ref{asymmfscal}).
The non-trivial critical point at $\sigma_c>0$, appearing by low diffusion 
rates exhibits the universal behavior of the transition of the 
$2A\to 3A$, $2A\to\emptyset$ (PCPD) model owing to the generation of 
the effective $2A\to\emptyset$ reaction via the quick processes: 
$2A\to 3A\to 4A\to\emptyset$ \cite{Ccom}.

Very recent studies by Canet et al. \cite{CPRL,CCD04} found similar diffusion
dependence in the $A\to 2A$, $2A\to\emptyset$ BARW model. 
Using non-perturbative renormalization group method (NPRG) they obtained 
a different phase diagram than what was expected by perturbative 
renormalization (RG) arguments \cite{Cardy-Tauber}. 
Namely directed percolation (DP) transition were found in these
models for $d>2$ dimensions with $d_c=4$.
Furthermore those transitions at $\sigma_c>0$ branching rate
persist up to infinite dimensions provided $\lambda/D$ is greater than
a threshold value. This threshold value was found to be zero for $d<3$
and finite for $d\ge 3$. 

In this work I apply dynamical, cluster mean-field approximations for
that BARW model and by extrapolations to $N\to\infty$ cluster sizes 
I determine the phase diagram in 1d. 
Furthermore I explore the phase transitions of the $A\to 2A$, $4A\to\emptyset$ 
model with this method and by simulations in one and two dimensions.
\begin{figure}
\begin{center}
\epsfxsize=87mm
\centerline{\epsffile{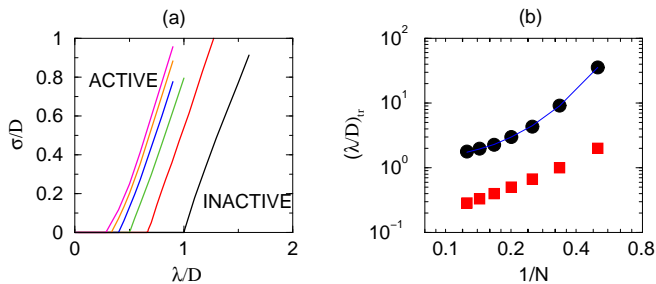}}
\caption{(a) Phase diagram of the $A\to 2A$, $2A\to\emptyset$ model
determined by $N=3,4,..8$ (right to left curves) cluster approximations.
(b) The $(\lambda_c/D)_{tr}$ critical endpoint values of the $\sigma_c>0$ 
transitions as the function of $1/N$.
Boxes correspond to the $A\to 2A$, $2A\to\emptyset$, while circles to the 
$A\to 2A$, $4A\to\emptyset$ model. The line shows a fitting of the
form (\ref{fitform})}
\label{fig1}
\end{center}
\end{figure}
The cluster mean-field method applied for nonequilibrium models first by
\cite{gut87,dic88} are based on the calculation of $N$-block probabilities 
of the model. This method has successfully predicted the phase
diagram of many systems. I apply it for one dimensional, site restricted 
lattice versions of the models mentioned before. It is well known that such 
approximations predict the phase structure qualitatively well in 1d. 
It provides an efficient phase diagram exploration method and could give 
valid results in higher dimensions too (see example \cite{bin2dcikk}).
One can set up a master equations for the  $P_N$ block probabilities as
\begin{equation}
\frac{\partial P_N(\{s_i\})}{\partial t} = f\left (P_N(\{s_i\})\right) \ ,
\label{mastereq}
\end{equation}
where the site variables take the values: $s_i=\emptyset,A$.
During the solution of these equations one estimates larger than $N$ 
sized block probabilities by the maximum overlap approximation like
\begin{equation}
P_{N+1}(s_1,...s_{N+1}) \simeq 
\frac{P_N(s_1,...s_{N}) P_N(s_2,...s_{N+1})}
{P_N(s_2,...s_{N},\emptyset)+P_N(s_2,...s_{N},A)}  \ . 
\label{approxeq}
\end{equation}
Taking into account spatial symmetries, for the maximal $N=8$ approximations
of this work one has to find the solution of equations of 136 independent 
variables. Using this method first I investigated the model defined by the 
transition probabilities
\begin{equation}
A \stackrel{\sigma}{\to} 2A,
\qquad 2A \stackrel{\lambda}{\to} \emptyset,
\qquad\emptyset A\stackrel{D}{\leftrightarrow} A\emptyset \ .
\label{1220rates}
\end{equation}
The steady state solutions have been determined for $N=1,2,3...,8$ 
approximations and the corresponding steady state densities 
$\rho_s(\sigma,D)$ are calculated numerically. The phase transition points
are obtained for several values and are plotted in the $\sigma_c/D$ vs.
$\lambda_c/D$ parameterization on plot (Fig.\ref{fig1}(a))
in order to be comparable with the results of \cite{CCD04}. The phase 
transition lines for $N>1$ corroborate the phase diagram of \cite{CCD04}, 
with $(\lambda_c/D)_{th}>0$ threshold values and with linear shapes in 
the $\sigma/D \to \infty$ limit. 
By inspecting the numerical values of the $(\lambda_c/D)_{th}(N)$ 
thresholds values for $N=2,3...8$ (see Table \ref{tab}) 
one can setup a hypothesis that in general
\begin{equation}
(\lambda_c/D)_{th}(N) = 2/(N-1)
\label{1220fitform}
\end{equation}
exactly. Therefore one can extrapolate that in the $N\to\infty$ limit
$(\lambda_c/D)_{th}(N) \to 0$ (Fig.\ref{fig1}(b)) in agreement with the 
results of \cite{CCD04} for 1d. So $N>1$ cluster mean-field approximations, 
which can take into account the diffusion exhibit similar phase diagram as 
the one obtained by the NPRG method and in the asymptotic limit they even 
reproduce the zero threshold corresponding to 1d.
\begin{figure}
\begin{center}
\epsfxsize=60mm
\centerline{\epsffile{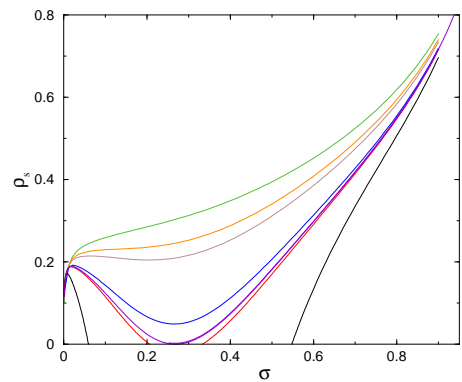}}
\caption{Steady state density results in the $N=7$ approximations of
the $A\to 2A$,  $4A\to\emptyset$ model. Different curves correspond to:
$D=1$, $0.7$, $0.6$, $0.4$, $0.371$, $0.36$, $0.2$ (top-to bottom). 
The $\sigma_c>0$ critical point disappears for $D^* > 0.37083$.}
\label{ou7}
\end{center}
\end{figure}

The other system of interest in this study is the
\begin{equation}
A \stackrel{\sigma}{\to} 2A,
\qquad 4A \stackrel{\lambda}{\to} \emptyset,
\qquad\emptyset A\stackrel{D}{\leftrightarrow} A\emptyset
\label{1240rates}
\end{equation}
model, for which perturbative RG predicts $d_c=2/3$ \cite{Cardy-Tauber}, 
therefore in all physical dimensions only type (\ref{asymmfscal}) 
mean-field transitions are expected at $\sigma_c=0$. By solving the
cluster mean-field equations up to $N=8$, with $\lambda=1-\sigma$
parameterization one obtains a reentrant phase structure for low
diffusion rates as in case of the $2A\to 3A$, $4A\to\emptyset$ model
\cite{2340cikk} (Fig.\ref{ou7}). I have determined the $(\lambda_c/D)_{th}$ 
threshold values for each $N$ and plotted them as the function of $1/N$ 
(Fig.\ref{fig1}(b)). One can see a tendency towards level-off 
contrary to the results for the $A\to 2A$, $2A\to\emptyset$ case.
By extrapolating to the $N\to\infty$ limit using a form similar, but
more general than (\ref{1220fitform})
\begin{equation}
(\lambda_c/D)_{th}(N) = a + b/(N-1)^c
\label{fitform}
\end{equation}
the threshold does not vanishes : $a=1.2$, $b=34.5$, $c=2.14$,
therefore we may expect phase transitions with $\sigma_c>0$ 
even in one dimension for any $0<D\le 1$ diffusion probabilities.
\begin{figure}
\begin{center}
\epsfxsize=60mm
\centerline{\epsffile{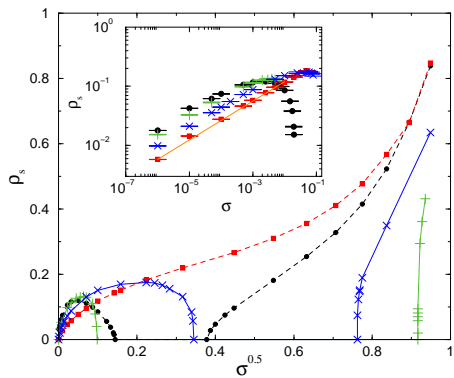}}
\caption{Simulation results for the steady state density of the
$A\to 2A$, $4A\to\emptyset$ model. Crosses correspond to $D=1.0$,
``+'' signs to $D=0.2$ diffusions in 1d. Circles denote $D=0.01$, squares
$D=1$ data in 2d. Error bars are smaller than the symbol sizes.
The lines serve to guide the eye. The insert shows the data magnified 
in the neighborhood of the $\sigma=0$ 
transition point. The solid line shows a power-law fitting with the
exponent $\beta=0.33(1)$.}
\label{comb1d2d}
\end{center}
\end{figure}
To test the cluster mean-field results for the $A\to 2A$, $4A\to\emptyset$ 
model I performed simulations in one dimension with lattice sizes: $L=10^5$
and with periodic boundary conditions. The simulations were started from
homogeneously filled lattices with probability 1/2. 
One elementary Monte Carlo step consist of the following sub-steps. 
A particle and a direction are selected randomly. A particle hopping is
attempted in the given direction with probability $D$ provided the nearest
neighbour (nn) site was empty. If the diffusion attempt was unsuccessful
either 4 nn particles in the given direction are removed with probability 
$\lambda$ or an offspring is created in an empty nn site in the direction
selected before. The time ($t$) measured by Monte Carlo steps (MCS)
is updated by $\Delta t = 1/n$, where $n$ is the total number of particles. 
The density of particles $\rho(t)$ is followed up to $10^7$ MCS 
(throughout the whole paper the time is measured by Monte Carlo steps (MCS)).
The phase transition points have been located for several diffusion rates 
and are shown on Fig.\ref{comb1d2d}. A qualitative agreement with the low-$D$ 
results of the finite $N$-cluster mean-field approximations can be observed.
However the $\sigma_c>0$ transition persists even for $D=1$ diffusion 
probability in agreement with the $N\to\infty$ extrapolations.

\begin{figure}
\epsfxsize=82mm
\epsffile{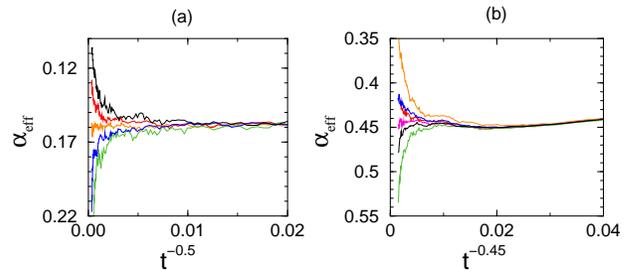}
\caption{(a) Local slopes of the density decay exponent $\alpha$ as the 
function of $t^{-0.5}$ of the  the one-dimensional $A\to 2A$, $4A\to\emptyset$ 
model at $D=0.2$. Different curves correspond to $\sigma=0.8399$, $0.83985$, 
$0.8398$, $0.83975$, $0.8397$ (top to bottom). The critical point is located
at $\sigma_c=0.83980(2)$. (b) The same as in (a) in two dimensions, for
$\sigma=0.1426$, $0.14262$, $0.14263$, $0.14264$, $0.14265$, $0.1427$
(top to bottom).}
\label{12401d}
\end{figure}

The density decay near a $\sigma_c>0$ critical point, at $D=0.2$ has been 
investigated in more detail. 
The density at $\sigma_c$ is expected to decay as (\ref{alphascale}). 
Fig.\ref{12401d}(a) shows the local slopes of $\alpha$ defined as
\begin{equation}
\alpha_{eff}(t) = {- \ln \left[ \rho(t) / \rho(t/m) \right] 
\over \ln(m)} \ , \label{slopes}
\end{equation}
(where I used $m=2$). As one can see curves with $\sigma>0.8398$ veer up
(active phase), while those with $\sigma<0.8398$ veer down (absorbing phase).
A clean scaling with $\alpha=0.159(1)$ exponent can be observed at  
$\sigma=0.83980(2)$ corroborating the 1+1 dimensional DP class 
exponent value ($\alpha=0.159464(6)$ \cite{Jensen99a}). In the inactive phase
$\rho(t)$ vanishes exponentially in agreement with the DP behavior again.
One may expect the same kind of transition all along the $\sigma_c(D) > 0$ 
transition line. Indeed simulations showed that the density decays in a 
similar way at transition points with $D=0.01$, $0.05$, $0.09$.
The simulations for this model were repeated in two dimensions in 
$L=4\times 10^3$ linear sized systems with periodic boundary conditions. 
These simulations were started from fully occupied lattices.
One elementary Monte Carlo step consists of the following processes.
A particle and a number $x_1 \in (0,1)$ are selected randomly; 
if $x_1 < D$ a site exchange is attempted with one of 
the randomly selected nn. The time is updated by $1/n$.
A particle and a number $x_2 \in (0,1)$ are selected randomly.
If $x_2 < \sigma$ and if the number of nn empty sites is greater than 0,
one new particle is created at an empty site selected randomly.
If $x_2 \ge \sigma$ and the number of nn particles is greater than 2,
four randomly selected neighboring particles are removed.
The time ($t$) is updated by $1/n$ again.
The density of particles was followed up to $t_{max} \le 10^7$ MCS.
As one can see simulation data (Fig.\ref{comb1d2d}) and the 7-point 
approximations (Fig.\ref{ou7}) fit qualitatively well (this is true
for other $N>1$ levels as well).
In both cases for weak diffusion rates reentrant phase transitions occur 
with $\sigma_c > 0$, while for strong diffusions only a single phase 
transition at $\sigma_c=0$ branching rate can be found. 
The density decay at $D=0.01$ is analyzed near the phase transition point.
The local slope figure shows a separatrix for the critical value
$\sigma_c=0.14263(1)$ as can be seen on Fig.\ref{12401d}(b). 
One can read-off the corresponding decay exponent $\alpha=0.445(5)$, 
which agrees with the 2+1 dimensional DP value 
($\alpha=0.4505(10)$ \cite{VoigtZiff97}).

I also investigated the steady state behavior at the $\sigma_c=0$ transition. 
The steady state density in the active phase near the critical phase 
transition point is expected to scale as eq.(\ref{betascale}).
As the insert of Fig.\ref{comb1d2d} shows by applying power-law fitting in 
the $10^{-6} \le\sigma\le 10^{-3}$ region one obtains $\beta \simeq 0.33(1)$
both in one and two dimensions at different diffusion rates. 
This agrees with the mean-field value (\ref{asymmfscal}) for this model. 
The density at $\sigma_c=0$ decays as $\rho\propto t^{-1/3}$ trivially, 
dictated by the $4A\to\emptyset$ process.

In conclusion numerical evidence is provided that the $A\to 2A$, 
$4A\to\emptyset$ branching and annihilating random walk process 
exhibits diffusion dependent phase transitions contrary the 
expectations coming from perturbative renormalization.
In particular $N$-cluster mean-field approximations (with extrapolations
to the $N\to\infty$ limit) resulted in a reentrant ($D$-$\sigma$) phase 
diagram, with  phase transitions at $\sigma_c>0$ for $0<D\le 1$ diffusion 
probabilities. Simulations have shown that along
this transition line DP critical behavior occurs in one and two 
dimensions. This type of critical behavior is the consequence of
an effective $A\to\emptyset$ reaction generated by  
$A\to 2A \to 3A\to 4A\to\emptyset$ of slowly moving particles.
In one dimension this line of phase transitions 
persist for all $0<D\le 1$ diffusions, while in 2d it disappears for high 
diffusion rates. For any $D$ and $d$ values a mean-field transition, 
characterized by $\beta=1/3$ occurs at $\sigma_c=0$.

Similar reentrant phase diagram has been observed in case of the one
dimensional $A\to 2A$, $3A\to\emptyset$ model \cite{Dicktrip},
in a variant of the NEKIM model \cite{nbarw2cikk} and in the 
$2A\to 3A$, $4A\to\emptyset$ model \cite{2340cikk,bin2dcikk}.
In all cases the diffusion competes with particle reaction processes,
and the bare parameters should somehow form renormalized reaction rates
which govern the evolution over long times and distances.
An other study using exact methods \cite{PS02} showed that the 
particle density fluctuations undergo a diffusion dependent phase transition
in the bosonic PCPD model for $d>2$. 
A very recent, non-perturbative RG study \cite{CCD04} has found similar 
diffusion dependent phase diagram in the $A\to 2A$, $2A\to\emptyset$ 
model. That work points out that non-perturbative effects arise
and there is a threshold $(\lambda/D)_{th}(d)$ above which DP, while below 
it a type (\ref{asymmfscal}) mean-field transition at $\sigma_c=0$ appears.
The present $N$-cluster approximations confirm those results and
eventuate similar phase diagram as in \cite{CCD04} for any finite $N$, 
with a threshold value: $(\lambda_c/D)_{th}(N) = 2/(N-1)$. 
In the $N\to\infty$ limit this corroborates a vanishing threshold in
1d \cite{CCD04}.

{\bf Acknowledgements:}

The author thanks I. Georgiev, G. Sch\"utz and U. T\"auber for the useful 
comments. The author thanks the access to the NIIFI Cluster-GRID, LCG-GRID 
and to the Supercomputer Center of Hungary.
Support from the Hungarian research fund OTKA (Grant No. T-046129)
is acknowledged.

\begin{table}[h]
\begin{center}
\begin{tabular}{|c|c|c|c|c|c|c|c|}
\hline
$N$                   &     2 & 3&     4 &   5 &    6 &    7 &     8\\
\hline
$(\lambda_c/D)_{th}$  &     2 & 1&  0.666& 0.5 &  0.4 & 0.333& 0.286 \\
\hline
\end{tabular}
\vskip 1mm
\caption{Numerical N-cluster results for the threshold values of 
the $A\to 2A$, $2A\to\emptyset$ model}.
\label{tab}
\end{center}
\end{table}
\end{multicols}

\begin{references}
\bibitem{DickMar} J.~Marro and R.~Dickman,
\newblock {\em Nonequilibrium phase transitions in lattice models},
\newblock Cambridge University Press, Cambridge, 1999.
\bibitem{Berry} H. Berry, Phys. Rev. E {\bf 67}, 031907 (2003).
\bibitem{Bou} J.-P.Bouchaud, A. Georges, Phys. Rep. {\bf 195}, 127 (1990).
\bibitem{Hin2000} H.~Hinrichsen, Adv. Phys. {\bf 49}, 815 (2000)
\bibitem{Uweof} U. T\"auber, Adv. in Solid State Phys. {\bf 43} 659 (2003).
\bibitem{dok} G. \'Odor, Rev. Mod. Phys. {\bf 76} (2004) in press, eprint:
cond-mat/0205644.
\bibitem{mexprocikk}G. \'Odor and N. Menyh\'ard, 
Physica D {\bf 168}, 305 (2002).
\bibitem{PCP} I. Jensen and R. Dickman, Phys. Rev. E {\bf 48}, 1710 (1993);
I. Jensen, Phys. Rev. Lett. {\bf 70}, 1465 (1993).
\bibitem{HaOd98} H. Hinrichsen and G. \'Odor, 
Phys. Rev. E {\bf 58}, 311 (1998).
\bibitem{tripcikk} G. \'Odor, Phys. Rev. E {\bf 67}, 056114 (2003).
\bibitem{PHK02} K. Park, H. Hinrichsen and I. Kim, Phys. Rev. E {\bf 66},
025101 (2002).
\bibitem{2340cikk} G. \'Odor, Phys. Rev. E {\bf 69}, 036112 (2004).
\bibitem{bin2dcikk} G. \'Odor, cond-mat/0403562. 
\bibitem{Ccom} Confirmed by H. Chat\'e in a private communication.
\bibitem{CPRL} L. Canet, B. Delamotte, O. Deloubrière and N. Wschebor,
Phys. Rev. Lett. {\bf 92}, 195703 (2004). 
\bibitem{CCD04} L. Canet, H. Chat\'e and B. Delamotte, cond-mat/0403423.
\bibitem{Cardy-Tauber} J. L. Cardy and U. C. T\"auber,
J. Stat. Phys. {\bf 90}, 1 (1998).
\bibitem{gut87} H. A. Gutowitz, J. D. Victor and B. W. Knight,  
Physica {\bf 28D}, 18, (1987).
\bibitem{dic88} R. Dickman, Phys. Rev. A {\bf 38}, 2588, (1988).
\bibitem{Jensen99a} I.~Jensen, J. Phys. A {\bf 32}, 5233 (1999).
\bibitem{VoigtZiff97} C.~A. Voigt and R.~M. Ziff, Phys. Rev. E {\bf 56},
R6241 (1997).
\bibitem{Dicktrip} R. Dickman, Phys. Rev. B {\bf 40}, 7005 (1989);
R. Dickman, Phys. Rev. A {\bf 42}, 6985 (1990).
\bibitem{nbarw2cikk} N. Menyh\'ard and G. \'Odor, 
Phys. Rev. E {\bf 68}, (2003) 056106.
\bibitem{PS02} M. Paessens, G. M. Schuetz, J. Phys. A {\bf 37} (2004) 4709.
\end{references}
\end{document}